\documentclass[fleqn,usenatbib,useAMS]{mnras}
\usepackage{newtxtext,newtxmath}
\usepackage[T1]{fontenc}
\usepackage{amsmath}
\DeclareRobustCommand{\VAN}[3]{#2}
\let\VANthebibliography\thebibliography
\def\thebibliography{\DeclareRobustCommand{\VAN}[3]{##3}\VANthebibliography}

\usepackage[dvipdfmx]{graphicx}	
\usepackage{amsmath}	
\usepackage{color}

\newcommand{\msun}{M_\odot} 
\newcommand{\rsun}{R_{\rm sun}} 
\newcommand{\yr}{\rm yr} 
\newcommand{\Gyr}{\rm Gyr} 
\newcommand{\pc}{\rm pc} 
\newcommand{\kms}{\rm km~s^{-1}} 
\newcommand{\kelvin}{\rm K} 
\newcommand{\gauss}{\rm G}

\newcommand{\mns}{M_{\rm NS}} 
\newcommand{\rns}{R_{\rm NS}} 
\newcommand{\robs}{R_{\rm obs}} 
\newcommand{\rmax}{R_{\rm max}} 
\newcommand{\mlv}{m_{\rm l,g}} 
\newcommand{\mnsv}{M_{\rm NS,g}} 
\newcommand{\tns}{T_{\rm NS}} 
\newcommand{\vkick}{v_{\rm kick}} 
\newcommand{\tauns}{\tau_{\rm NS}} 

\newcommand{\sgmns}{\dot{\Sigma}_{\rm NS}} 
\newcommand{\sgmsfr}{\dot{\Sigma}_\ast} 
\newcommand{\fns}{f_{\rm NS}} 

\newcommand{\nns}{N_{\rm NS}} 
\newcommand{\dmax}{D_{\rm max}} 
\newcommand{\dmin}{D_{\rm min}} 
\newcommand{\taumax}{\tau_{\rm max}} 
\newcommand{\taumin}{\tau_{\rm min}} 

\title[Hunting neutron stars with wide-area optical surveys]{Hunting isolated neutron stars with proper motions from wide-area optical surveys}

\author[D.~Toyouchi et al.]{
Daisuke~Toyouchi$^{2,1}$\thanks{E-mail: toyouchi@resceu.s.u-tokyo.ac.jp},
Kenta~Hotokezaka$^{2,1}$,
and Masahiro~Takada$^{1}$
\\
$^{1}$Kavli Institute for the Physics and Mathematics of the Universe
(WPI), 
The University of Tokyo Institutes for Advanced Study (UTIAS),\\
The University of Tokyo, Chiba 277-8583, Japan\\
$^{2}$Research Center for the Early Universe (RESCEU), The University of Tokyo
Hongo, 7-3-1, Bunkyo-ku
Tokyo, 113-0033, Japan}

\date{Accepted XXX. Received YYY; in original form ZZZ}

\pubyear{2021}

\begin{document}
\label{firstpage}
\pagerange{\pageref{firstpage}--\pageref{lastpage}}
\maketitle

\begin{abstract}
High-velocity neutron stars (HVNSs) that were kicked out from their birth location can be potentially identified with their large proper motions, and possibly with large parallax, when they come across the solar neighborhood.
In this paper, we study the
feasibility of hunting isolated HVNSs in wide-area optical surveys by modeling the evolution of NS luminosity taking into account spin-down and thermal radiation. 
Assuming the upcoming 10-year VRO LSST observation,
our model calculations predict that  
about 10 HVNSs mainly consisting of pulsars with ages of $10^4$--$10^5~\yr$ and thermally emitting NSs with $10^5$--$10^6~\yr$
are detectable.
We find that a few NSs with effective temperature 
$< 5 \times 10^5~\kelvin$, which are likely missed in the current and future X-ray surveys, are also detectable.
In addition to the standard neutron star cooling models, we consider
a dark matter heating model.
 If such a strong heating exists we find that the detectable HVNSs would be significantly cooler, i.e., $\lesssim 5\times 10^5\,{\rm K}$.
Thus, the future optical observation will give an unique NS sample, which can provide essential constraints on the NS cooling and heating mechanisms.
Moreover, we suggest that providing HVNS samples with optical surveys is helpful for understanding the intrinsic kick-velocity distribution of NSs.
\end{abstract}

\begin{keywords}
astrometry -- stars: neutron -- pulsars: general
\end{keywords}



\section{Introduction}\label{sec:intro}

Neutron stars (NSs) have been generally observed as high-velocity objects with typical velocities of several hundred kilometer per second \citep[e.g.,][]{Lyne1994Natur, Hansen1997MNRAS, Cordes1998ApJ,Fryer1998ApJ,Walter2001ApJ,Kaplan2002ApJ,Walter2002ApJ,Hobbs2005MNRAS,Kaplan2007ApJ, Ronchi2021arXiv}, which are much larger than that of their presumed progenitors, i.e., massive OB stars, and other
majority of main-sequence stars.
A possible mechanism providing such high-velocity neutron stars (HVNSs) is anisotropic supernova explosion that kicks out the remnant core following the momentum conservation law \citep[e.g.,][]{Shklovskii1970SvA, Dewey1987ApJ,Burrows1995ApJ,Scheck2004PhRvL, Wongwathanarat2013A&A,Janka2017ApJ,Nagakura2017ApJS,Nakamura2019PASJ,Mandel2020MNRAS}. For low mass progenitors, however, the explosion is less energetic and less asymmetric, and thus, the hydrodynamical kicks are expected to be smaller \citep{Melson2015ApJ,Suwa2015MNRAS,Radice2017ApJ,Gessner2018ApJ}. In fact, extremely small kicks are required from the orbital parameters of binary neutron stars \citep{Piran2005PhRvL,Beniamini2016MNRAS,Tauris2017ApJ}.
In addition to the hydrodynamical kicks, the disruption of a close binary system and the anisotropic neutrino emission can also induce the kick velocities \citep[e.g.,][]{Gott1970ApJ, Iben1996ApJ,Lai1998ApJL}.
Thus, exploring the velocity distribution of HVNSs is an essential subject to study the physics of NS formation.

Measurements of NS proper motions have been mainly done with radio timing and interferometric observations \citep[][and references therein]{Lorimer2004hpa}, and therefore highly biased to radio-luminous pulsars. The discovery of  thermally emitting isolated NSs by {\it ROSAT} all-sky survey provided a unique opportunity to study the nature of isolated NSs including the thermal history of NSs and natal kicks \citep[e.g.,][]{Haberl2007Ap&SS}.
In fact, the optical counterparts of five of them were identified and the tangential velocities of  RX~J1856.5-3754 and RX~J0720.4-3125, are obtained as $\sim 200\,{\rm km\,s^{-1}}$ by using astrometric data from the {\it Hubble Space Telescope} \citep[][]{Walter2001ApJ,Kaplan2002ApJ,Walter2002ApJ,Kaplan2007ApJ}. Note that the proper motions of some of the known thermal emitting isolated NSs are also measured by ground-based optical telescopes and the Chandra Observatory \citep[e.g.,][]{Motch2003A&A,Motch2005A&A,Zane2006A&A,Motch2009A&A}. 
We envision
that the upcoming {\it eROSITA} all-sky survey
\citep{Sunyaev2021}
will 
significantly increase the number of known thermally emitting isolated NSs \citep{Pires2017AN} 

Optical identifications of NSs have been still limited because they are extremely faint in 
optical bands. 
However, future deep optical surveys, such as the Vera Rubin Observatory's LSST\footnote{\url{https://www.lsst.org}}, will give independent HVNS samples, characterized as faint and blue objects with large proper motions.
Providing optically selected NS samples is an important task in not only clarifying the velocity distribution, but also studying the
unexplored nature cold NSs with surface temperature of $\sim 10^{5}\,{\rm K}$ as well as the radiative mechanism powering optically emittting magnetars and pulsars.
In this paper, we quantitatively investigate the feasibility of hunting HVNSs in wide-area optical 
surveys and discuss how the future observations can be used to constrain the formation and evolution of NSs.

The rest of the paper is organized as follows. 
We first model the optical luminosity of NSs in Section~\ref{sec:model} and describe the method to quantify the detectability of HVNSs in Section~\ref{sec:cal}.
Based on the model calculations, we present the prediction of the number of detectable HVNSs and clarify the properties of the obtained solution in Section~\ref{sec:basic}, and also discuss the dependence of our results on several model parameters in Section~\ref{sec:dependence}.
Finally, the summary and conclusion are given in Section~\ref{sec:summary}.
















\section{Method}\label{sec:method}

\subsection{Modeling Optical Luminosity of Neutron Stars}\label{sec:model}

In this paper we consider 
the
possibility of finding neutron stars (NS) with an optical observation. 
An NS is extremely faint due to its small size ($\rns \sim 10~$km for the radius), but it is not impossible to find if an NS exists in the solar neighborhood, say within a few 100~pc from the Earth. 
To make a quantitative estimate of detectability of NSs with an optical observation, we first need to model a luminosity of NS in optical wavelengths as a function of age. 

\begin{figure}
\begin{center}
\includegraphics[width=\columnwidth]{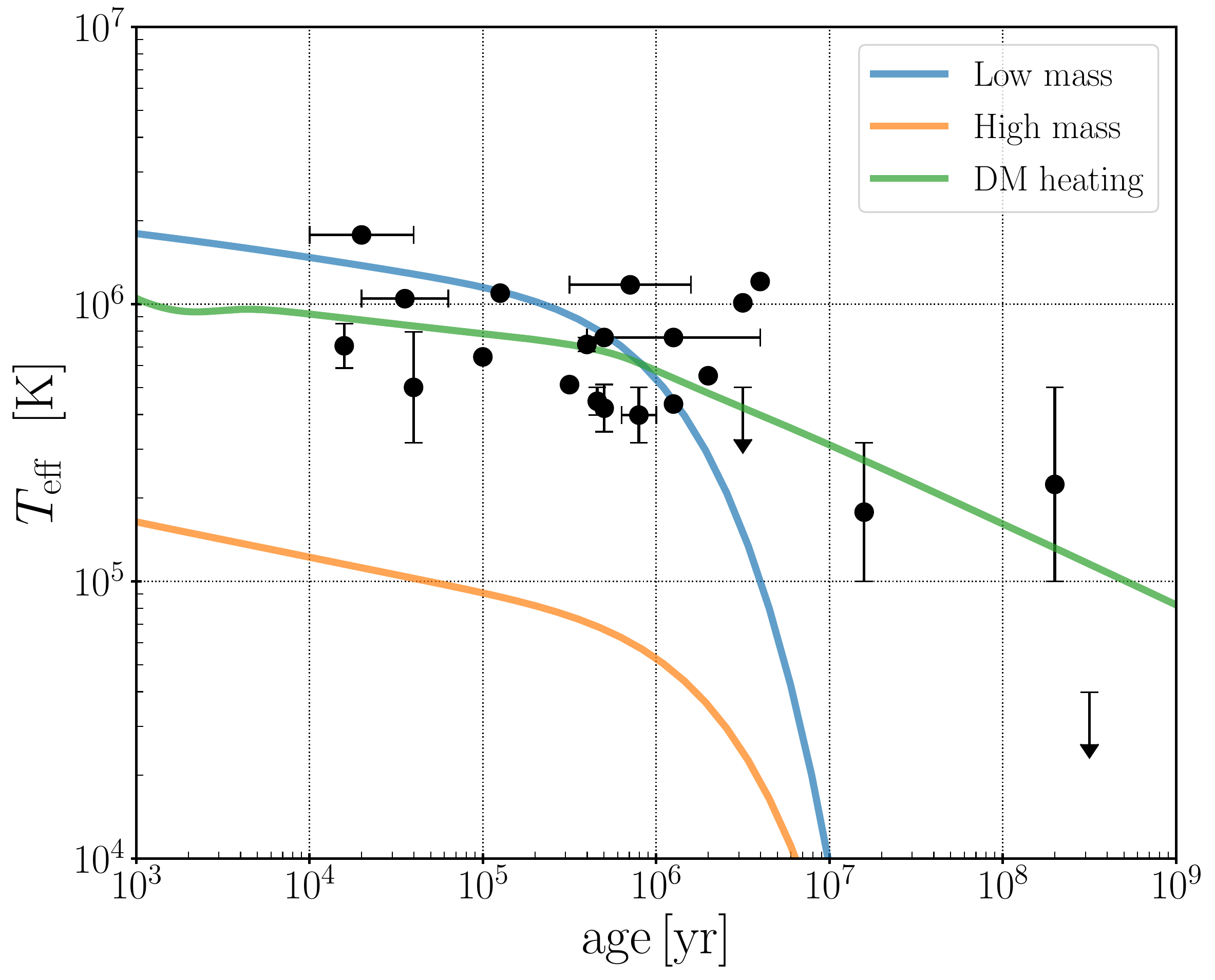}
\caption{Three NS cooling curves considered in this paper. The cyan, orange, and green lines represent Low-mass, High-mass, and DM heating models, respectively. For comparison, the observational data are overplotted \citep[see][for references]{Yanagi2020MNRAS}.}
\label{fig:cool_model}
\end{center}
\end{figure}

A radiation spectrum of NSs would generally consist of two contributions: 
thermal radiation emitted from the neutron surface and non-thermal radiation from magnetosphere.
Each contribution changes with time.
We assume that the thermal emission from the surface consists of
a single black body spectrum of an effective temperature, $\tns$, which declines with age of NSs, $\tauns$, assuming the energy loss via neutrino emission \citep[see, e.g., ][for details]{Tsuruta2002ApJ,Page2004ApJS,Potekhin2020MNRAS}.
Since the cooling history of NSs is still uncertain,
we here examine the uncertainty in our calculation by considering three NS cooling curves as displayed in Figure~\ref{fig:cool_model}.
Here, ``Low-'' and ``High-mass'' models are taken from Figure~11 of \citet{2004ARA&A..42..169Y}, and dark matter (DM) heating model is taken from
\citet{Yanagi2020MNRAS}.
Low-mass model assumes an NS mass of $\mns = 1.35~\msun$, in which the core density is too small to cause neutrino emission efficiently.
This cooling curve is generally in good agreement with the observed temperature of young NSs with $\tauns \lesssim 10^6$ yr.
Note here that the absence of NSs with $\tns \lesssim 5 \times 10^5~\kelvin$ would reflect the fact that the observed thermal flux is quickly dimmed in X-ray band, so that optical observations are advantageous in exploring the low-temperature regime.
To argue the detectability of such low-temperature NSs, we adopt High-mass model corresponding to greater NS mass case with $\mns = 1.98~\msun$, in which the thermal energy can be efficiently released via direct Urca process \citep{1995SSRv...74..455N}.

Note that both Low and High-mass models predict rapid decrements of $\tns$ at $> 10^6$ yr, in contrast to the thermal emission reported for very old millisecond pulsars \citep{Kargaltsev2004ApJ,Webb2019A&A}.
Several possibilities for the internal heating mechanism as well as the heating due to non-thermal particles in the magnetosphere are suggested \citep[e.g.][and references therein]{Gonzalez2010A&A,Yanagi2020MNRAS}.
Motivated by these, here we consider a dark matter (DM) and rotochemical  heating model, in which
the surface temperature of NSs stays $\gtrsim 10^{5}\,{\rm K}$ for $\sim 1\,{\rm Gyr}$ \citep{Hamaguchi2019PhLB}.
Including such heating sources will enhance the detectability of thermally emitting isolated NSs in optical wavelengths. 

We also calculate the non-thermal radiation that arises from the decay of
the spin and the magnetic field 
of NSs using an analytic model presented by \citet{2019MNRAS.487.1426B}, which successfully reproduces the observed 
distributions of
the spin period $P$ and the derivative $\dot{P}$ for known magnetars.
In this model, satisfying the general properties of spin-down, $\dot{\Omega} \propto B^2 \Omega^n$, and the magnetic field decay, $\dot{B} \propto B^{1+\alpha}$ \citep{2000ApJ...529L..29C}, the magnetic field and spin are assumed to evolve with time as
\begin{align}
B = B_0 \left ( 1 + \frac{\alpha t}{\tau_B} \right )^{-1/\alpha} \ ,
\label{eq:B_t}
\end{align}
and
\begin{align}
\Omega = \Omega_0 \left[ \frac{(n-1) \tau_B}{2(\alpha-2)\tau_\Omega} \left\{ \left (1 + \frac{\alpha t}{\tau_B} \right)^{\frac{1}{1-n}} -1 \right\} +1 \right ]^{\frac{1}{1-n}} \ ,
\label{eq:Omega_t}
\end{align}
where we set $\alpha = -1$ and $n = 3$ as adopted in \citet{2019MNRAS.487.1426B}, quantities with 
subscript 0 denote the 
quantities at the NS birth, 
and
 $\tau_B$ and $\tau_\Omega$ are the decay timescales of magnetic field and spin, respectively. Assuming a dipole magnetic field, we describe the spin-down timescale as,
\begin{align}
\tau_\Omega = \left |\frac{\Omega_0}{2\dot{\Omega}_0} \right | = \frac{3c^2 I }{4 B^2_0 \rns^6 \Omega_0^2}  \ ,
\label{eq:tau_Omega}
\end{align}
where $c$ is the speed of light, and $I$ is the inertia moment of NS.

\begin{figure}
\begin{center}
\includegraphics[width=\columnwidth]{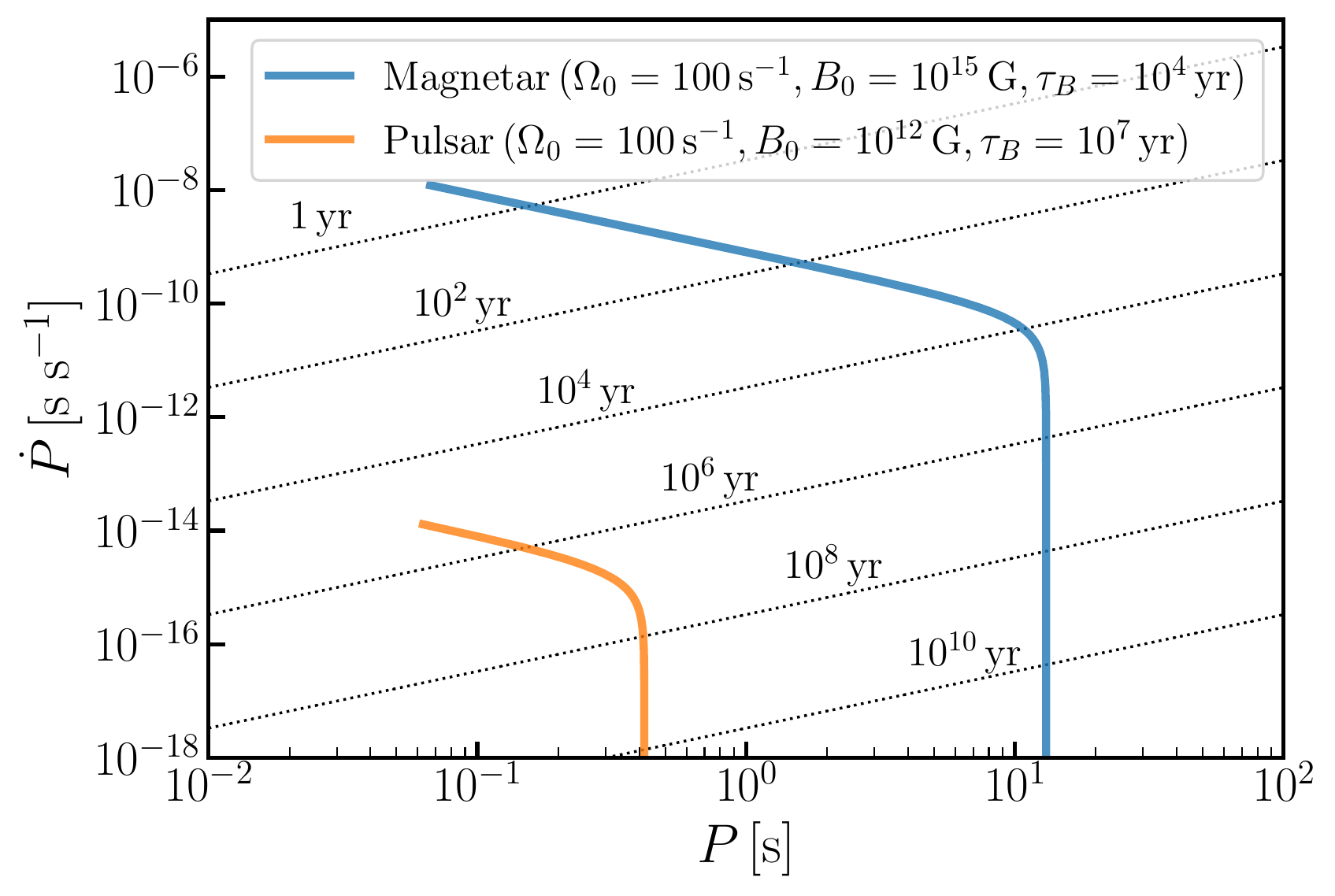}
\end{center}
\caption{Blue and orange lines show the spin evolution of 
magnetar-
and 
pulsar-like
NSs, respectively, calculated with the theoretical model by \citet{2019MNRAS.487.1426B}. Dotted lines represent the corresponding age of NSs calculated with $P/\dot{P}$.}
\label{fig:spindown}
\end{figure}

The spin-down evolution highly depends on the initial strength of magnetic field and its decay timescale.
In our calculation, we consider two NS models; a magnetar-like NS with $B_0 = 10^{15}\,\gauss$ and $\tau_B = 10^4$\,yr, and a pulsar-like NS with 
$B_0 = 10^{12}\,\gauss$ and $\tau_B = 10^7$\,yr,
and we set $\Omega_0 = 100\,{\rm s}^{-1}$ for both cases.
The spin-down evolution of two NS models is shown in Figure~\ref{fig:spindown}.
In general, the spin and magnetic field of NSs slowly declines when $\tauns \ll \tau_B$, and then suddenly decays
at $\tauns \sim \tau_B$.
The magnetar-like NSs have higher $\dot{P}$ at any $P$ and more rapidly lose their rotational and magnetic energies than the pulsar-like NSs do. 

\begin{figure}
\begin{center}
\includegraphics[width=\columnwidth]{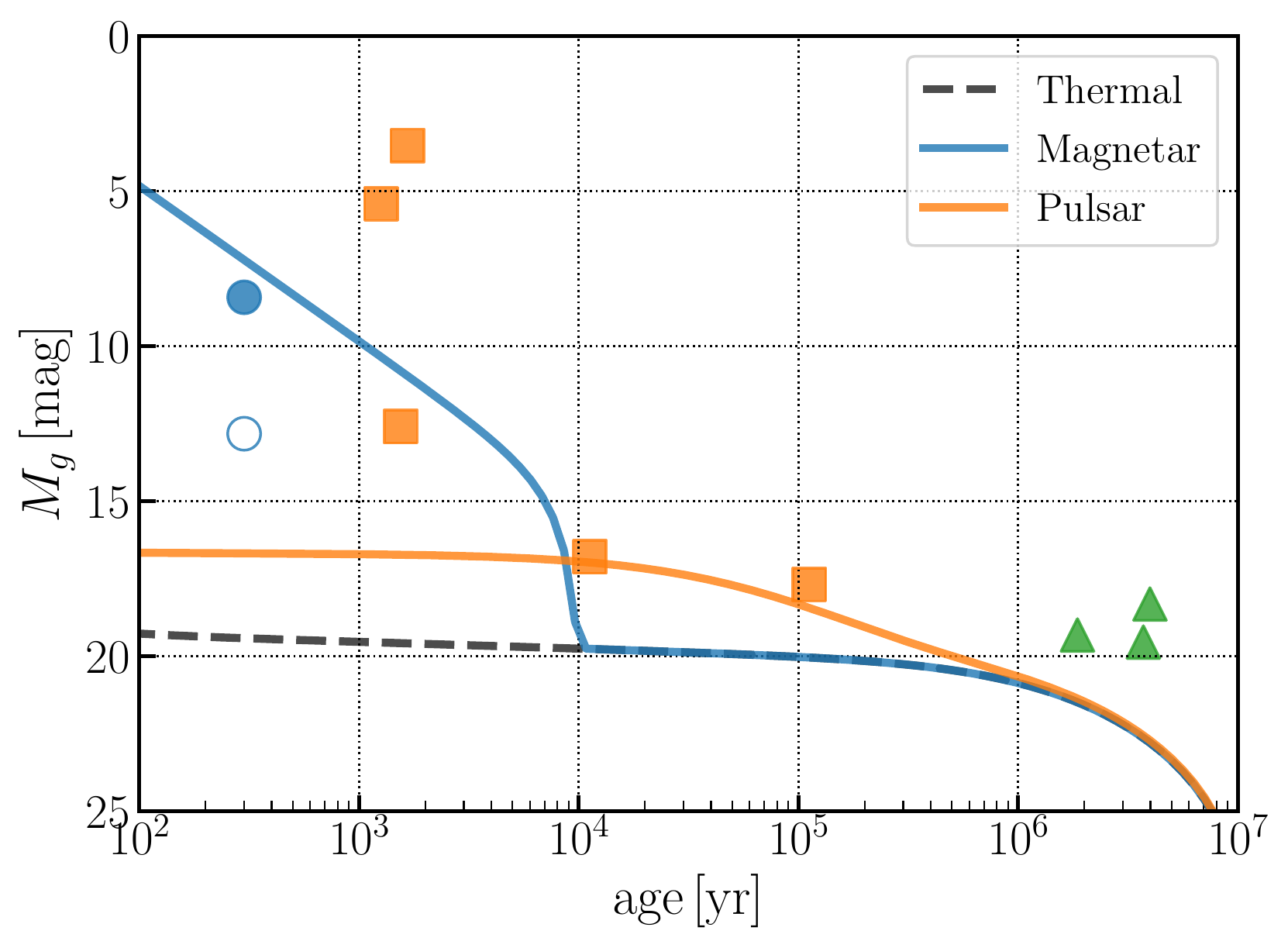}
\end{center}
\caption{Blue and orange lines represents the absolute magnitude in the $g$-band of magnetar- and pulsar-like NSs, respectively, calculated from the composite spectra of the spin-down and the black body radiation. The black dashed line denotes the optical luminosity only from the thermal radiation calculated with Low-mass cooling model (see Figure~\ref{fig:cool_model}). Blue circles, orange squares, and green triangles represent the observed magnetars, pulsars, and X-ray dim NSs, respectively. The plotted magnitudes are dust extinction corrected although the uncorrected magnitude is also indicated for the magnetar suffering from significant dust extinction.}
\label{fig:luminosity}
\end{figure}

With the spin-down energy loss, $\dot{E}_{\rm sd} = I \Omega \dot{\Omega}$, we assess the NS optical luminosity adopting the conversion factor of the spin-down energy to the optical radiation, $L_{\rm opt}/\dot{E}_{\rm sd} = 10^{-2}$ and $10^{-7}$ for magnetars and pulsars, respectively, based on the observational implication for isolated NSs \citep{Mereghetti2008A&AR,2009Msngr.138...19M,Enoto_2019}\footnote{Note that, in reality, the magnetars' optical luminosity may be proportional to $\dot{E}_{\rm mag}$ rather than the spin down luminosity, where $\dot{E}_{\rm mag}$ is the magnetic energy dissipation rate.
However, here we use the spin-down luminosity as a proxy for the optical luminosity for simplicity.}.
With this model
we consider the composite spectra of the spin-down and thermal radiation: 
Figure~\ref{fig:luminosity} shows 
the absolute 
magnitude for magnetar- and pulsar-like NSs
in optical wavelengths, compared to the observations of known NSs \citep{2009Msngr.138...19M}. 
While young magnetars with $\tauns < 10^4~\yr$ are brighter than $M_g \sim$ 15\,mag due to the greater spin-down radiation, 
the magnetars quickly become fainter after the magnetic field decays and then afterwards have a black body radiation
until $\tauns \sim 10^6$\,yr.
The spin-down radiation of pulsars is much weaker than the young magnetars, but the pulsars can be as bright as $M_g \sim 18\,$mag 
for $\sim 10^5$\,yr.
The optical luminosity of magnetars and pulsars at $\tauns > 10^6$\,yr is almost the same and 
arises from
the black body radiation, which fades with the NS cooling.
Note that our model 
is generally in good agreement with the observed optical luminosity of NSs except for young pulsars with $\sim 10^3~\yr$ that are thought to be extremely bright with the interaction between a central NS and the dense circum-stellar medium.
Since the radiation mechanism powering such young pulsars is poorly understood, in this paper we focus on the spin-down and thermal radiation and provide a conservative estimate in the number of detectable NSs.

\subsection{The Expected Number of HVNSs}\label{sec:cal}

Here we introduce our method to quantitatively calculate the detectability of HVNSs.
We assume that some of NSs have a large kick velocity, typically 500~${\rm km}~{\rm s}^{-1}$ \citep{Lyne1994Natur,Hobbs2005MNRAS}, and it would appear to have a large proper motion:
\begin{align}
\mu\simeq \frac{v_{\rm kick}}{d}\simeq 1.0~{\rm arcsec}~{\rm yr}^{-1}\left(\frac{v_{\rm kick}}{500~{\rm km}~{\rm s}^{-1}}\right)
\left(\frac{d}{100~{\rm pc}}\right)^{-1}.
\end{align}
%
If we have a 10-year monitoring observation of the sky such as the LSST, the NS
appears to have about 10~arcsec angular offset over 10~years, which should be easy
to identify if it is detected in optical wavelengths. 
In addition, an NS
in a close distance of 100~pc should have a relatively large parallax, $10~{\rm mas}$, which is not impossible to detect 
with a high-quality ground-based data \citep{2020arXiv200412899Q} or would be relatively easy to detect with a follow-up observation by an extremely large telescope such as the Thirty Meter Telescope (TMT) or
space-based observations such as the Roman Space Telescope. 
Hence, a task of finding an NS
with  an optical observation is to search for a very faint point-source object(s) that has a large proper motion, together with a possible 
detection of the parallax. 

\begin{figure}
\begin{center}
\includegraphics[width=\columnwidth]{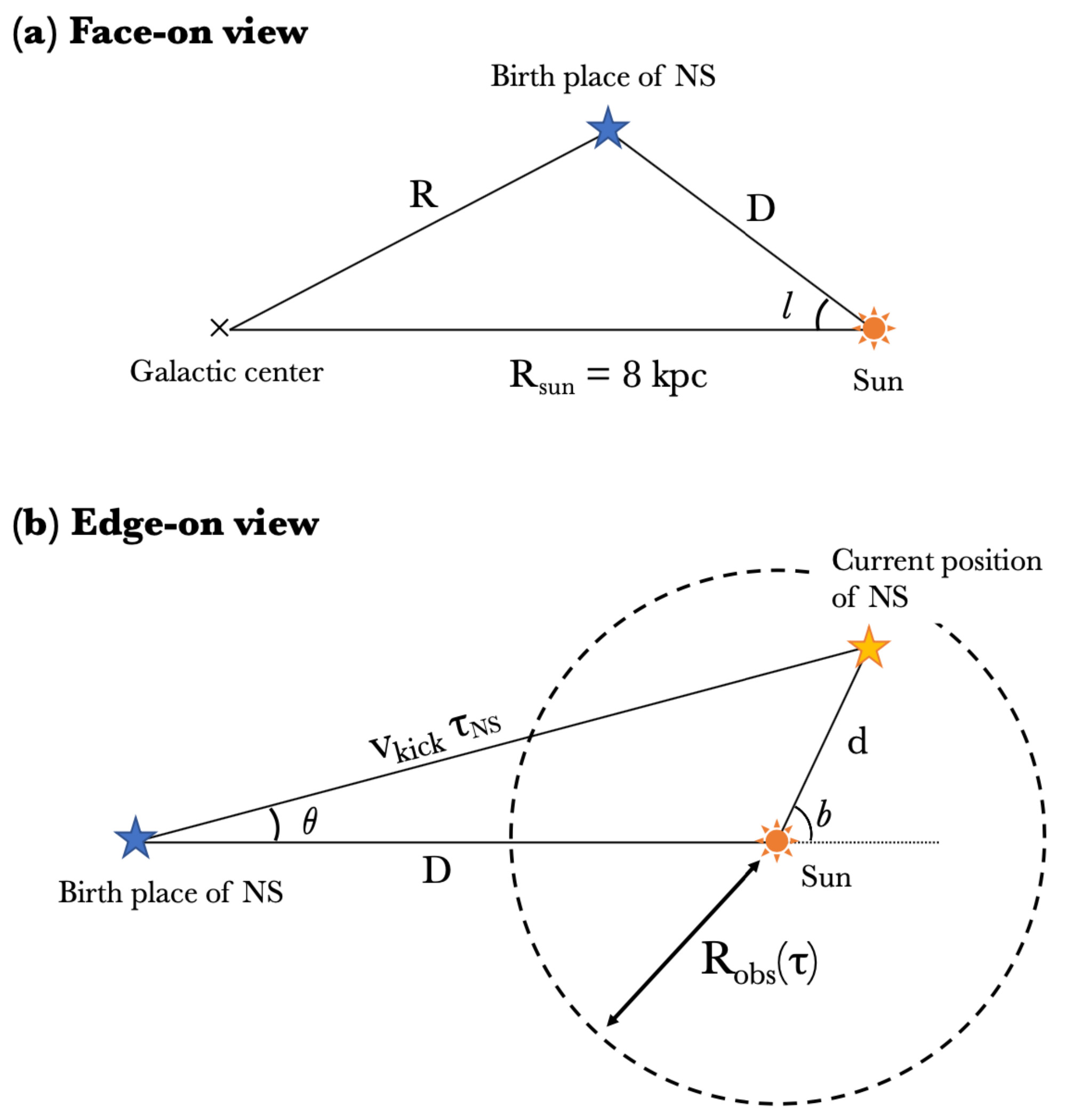}
\end{center}
\caption{The schematic picture shows the coordinate system that we adopt to calculate the number of detectable HVNSs. Here, the upper and lower pictures correspond to the face-on and edge-on views to the Galactic disk, respectively.}
\label{fig:coordinate}
\end{figure}

In order to estimate the expected number of detectable NSs
 in the solar neighborhood,
we consider the polar coordinate around the Sun as illustrated in Figure~\ref{fig:coordinate}.
We assume that each NS formed in the Galactic disk plane (i.e. ignore a possible offset of the newly formed NS in the vertical direction from the disk plane)
because NS progenitors are short-lived enough to be regarded as a member of kinematically cold thin disk stars. In this setting 
we assume that an NS formed at the place $(D, l)$ at birth and 
all NSs have a kick velocity of $\vkick = 500~\kms$ at birth with a random direction parametrized by the angle $\theta$ ($0\le \theta\le 2\pi$) in Figure~\ref{fig:coordinate}.
In this setting, 
the distance between the Sun and an NS with age of $\tauns$ can be given as
\begin{eqnarray}
d =  \sqrt{(\vkick\tauns)^2+D^2-2\vkick \tauns D{\rm cos}\theta} \ ,
\label{eq:rns}
\end{eqnarray}
and we consider NSs to be detectable if $d$ is sufficiently close to the Sun in that the luminosity is detectable: the observable radius is given as
\begin{eqnarray}
\robs = {\rm min} \left ( 10^{\frac{1+\mlv - M}{5}}~{\rm pc}, \ \rmax \right ) \ ,
\label{eq:robs}
\end{eqnarray}
with
\begin{eqnarray}
\rmax = 1~{\rm kpc}~\left( \frac{\vkick}{500~\kms} \right) \left(\frac{\mu_{\rm l}}{0.1~{\rm arcsec~yr^{-1}}} \right)^{-1} \ ,
\label{eq:rmax}
\end{eqnarray}
where $\mlv = 27~{\rm mag}$ is the $g$-band limiting magnitude in LSST observations.
The maximum radius $\rmax$ is set to detect NS proper motions, and in this work we suppose the LSST limiting proper motion, $\mu_{\rm l} \sim 0.1~{\rm arcsec~yr^{-1}}$.
As noted in Section \ref{sec:model}, the $g$-band absolute magnitude of NS $\mnsv$ declines with age of NSs, so that the observable radius can depends on age of NS, i.e., $\robs(\tauns)$.
Also note that the optical luminosity of magnetar- and pulsar-like NSs are much different as shown in Figure~\ref{fig:luminosity}.
We assume that a half of NSs are born as magnetars and the another half as pulsars, motivated by the observed number fraction of magnetars and pulsars  \citep{2019MNRAS.487.1426B}, and assess the detectability of each type of NSs.

By considering the maximum angle $\theta_{\rm max}$ for $d \leq \robs$, we describe the possibility that NS with an age of $\tauns$ launched at the distance $D$ from the Sun is coming into the observable regime with the solid angle $\Omega(\theta_{\rm max})$ as below,
\begin{eqnarray}
P(D, \tauns) &=& \frac{\Omega(\theta_{\rm max})}{4\pi} = \frac{1 - {\rm cos} \theta_{\rm max}}{2} \nonumber \\
&=& \frac{1}{2}-\frac{(\vkick \tauns)^2+D^2-\robs^2}{4 \vkick \tauns D} \ ,
\label{eq:prob}
\end{eqnarray}
where note that $0 \leq P \leq 1$.

We assume that HVNSs form in the disk region with a following time-independent rate,
\begin{eqnarray}
\sgmns = \fns \sgmsfr \ ,
\label{eq:sgmns}
\end{eqnarray}
with
\begin{eqnarray}
\sgmsfr = 3.5 \left ( \frac{R}{\rsun} \right )^2 {\rm exp} \left ( - 5 \frac{R-\rsun}{\rsun} \right )~\msun~\pc^{-2}~\Gyr^{-1} \ ,
\label{eq:sgmsfr}
\end{eqnarray}
where $\fns = 0.01$ is the number of NS per unit stellar mass, and $\sgmsfr$ is the radial profile of the star formation surface density rate in the MW presented by \cite{2015A&A...580A.126K}.
It is worth noting that this star-formation rate profile successfully represents the radial distribution of OB stars and radio pulsars along the MW disk \citep{1997ApJ...476..166W, 2006MNRAS.372..777L}, so that we practically trace the star formation history within about the last $10^7~\yr$.

With Eqs.~(\ref{eq:rns}-\ref{eq:sgmsfr}), we estimate the number of detectable HVNSs as below,
\begin{eqnarray}
\nns = \frac{1}{2} \int^{\dmax}_{\dmin} \int^{2\pi}_{0} \int^{\taumax}_{\taumin} \sgmns \left ( P_{\rm mag} + P_{\rm pul} \right ) {\rm d}D{\rm d}\phi{\rm d}\tauns \ ,
\label{eq:number}
\end{eqnarray}
where $P_{\rm mag}$ and $P_{\rm pul}$ are the observational probabilities calculated with Eq.~(\ref{eq:prob}) for the magnetar- and pulsar-like NSs, respectively.

\begin{table*}
\begin{center}
\caption{Model parameters and results for thermally emitting isolated neutron stars.} \label{table:model1}
\begin{tabular}{cccccccc} \hline \hline
 ID & Cooling model & $\vkick$ $[{\rm km~s^{-1}}]$ &   $N_{\rm tot}$ & $N_{\rm tot}(|b| > 5^\circ)$ & $N_{\rm tot}(T_{\rm eff} < 5 \times 10^5~\rm K)$ \\ \hline

1 & Low mass & 50 & 3.2 & 2.4 & 1.9 \\
2 & Low mass & 100 &   7.0 & 4.6 & 1.9 \\
3 & Low mass & 250 & 5.7 & 3.8 & 0.5 \\
4 & Low mass & 500 & 4.1 & 2.9 & 0.2 \\
5 & Low mass & 1000 & 2.7 & 2.0 & 0.2 \\ \hline
6 & High mass & 50 &  0.5 & 0.3 & 0.5 \\
7 & High mass & 100 &  0.3 & 0.2 & 0.3 \\
8 & High mass & 250 & 0.2 & 0.1 & 0.2 \\
9 & High mass & 500 &  0.1 & 0.1 & 0.1 \\
10 & High mass & 1000 & 0.1 & 0.1 & 0.1 \\ \hline
11 & DM Heating & 50 &  7.4 & 6.0 & 5.3 \\
12 & DM Heating & 100 & 11.6 & 8.8 & 5.0 \\
13 & DM Heating & 250 & 6.2 & 4.8 & 1.9 \\
14 & DM Heating & 500 & 3.8 & 2.9 & 0.8 \\
15 & DM Heating & 1000 &  2.2 & 1.7 & 0.3 \\ 
 \hline \hline
\end{tabular}
\end{center}
\end{table*}

\begin{table*}
\begin{center}
\caption{Model parameters and results for pulsars and magnetars.} \label{table:model2}
\begin{tabular}{ccccccc} \hline \hline
ID & Spin-down model & $\vkick$ $[{\rm km~s^{-1}}]$ & $\tau_{\rm B}$ [yr] &  $N_{\rm tot}$ & $N_{\rm tot}(|b| > 5^\circ)$ \\ \hline
16 & Pulsar & 50 & $10^{7}$ & 0.7 & 0.5  \\
17 & Pulsar & 100 & $10^{7}$ & 1.6 & 0.8  \\
18 & Pulsar & 250 & $10^{7}$  & 3.6 & 1.1  \\
19 & Pulsar & 500 & $10^{7}$ & 5.2 & 1.4  \\
20 & Pulsar & 1000 & $10^{7}$  & 4.9 & 1.7\\
21 & Pulsar & 500 & $10^{8}$ & 5.2 & 1.4  \\
22 & Pulsar & 500 & $10^{9}$  & 5.2 & 1.4  \\
 \hline
23 & Magnetar & 50 & $10^{4}$  & 0.0 & 0.0  \\
24 & Magnetar & 100 & $10^{4}$ & 0.0 & 0.0  \\
25 & Magnetar & 250 & $10^{4}$ & 0.1 & 0.0  \\
26 & Magnetar & 500 & $10^{4}$ & 0.4 & 0.0  \\
27 & Magnetar & 1000 & $10^{4}$ & 1.9 & 0.0  \\
 \hline \hline
\end{tabular}
\end{center}
\end{table*}

\begin{table*}
\begin{center}
\caption{Model parameters and results with composite spectra of the thermal and spin-down radiation (50\% pulsars and 50\% magnetars).} \label{table:model3}
\begin{tabular}{cccccccc} \hline \hline
ID & Cooling model & $\vkick$ $[{\rm km~s^{-1}}]$ &  $N_{\rm tot}$ & $N_{\rm tot}(|b| > 5^\circ)$ & $N_{\rm tot}(T_{\rm eff} < 5 \times 10^5~\rm K)$ \\ \hline
28 & Low mass & 100 & 7.3 & 5.0 & 2.1 \\
29 & Low mass & 500 & 10.0 & 4.3 & 0.2 \\ \hline
30 & High mass & 100 & 1.9 & 1.0 & 1.9 \\
31 & High mass & 500 & 5.8 & 1.5 & 5.8 \\ \hline
32 & DM Heating & 100 & 12.2 & 9.2 & 5.1 \\
33 & DM Heating & 500 & 9.7 & 4.4 & 0.8 \\
\hline \hline
\end{tabular}
\end{center}
\end{table*}


\section{Result}\label{sec:result}

\subsection{Basic Properties of the Solution}\label{sec:basic}

\begin{figure*}
\begin{center}
\includegraphics[width=2\columnwidth]{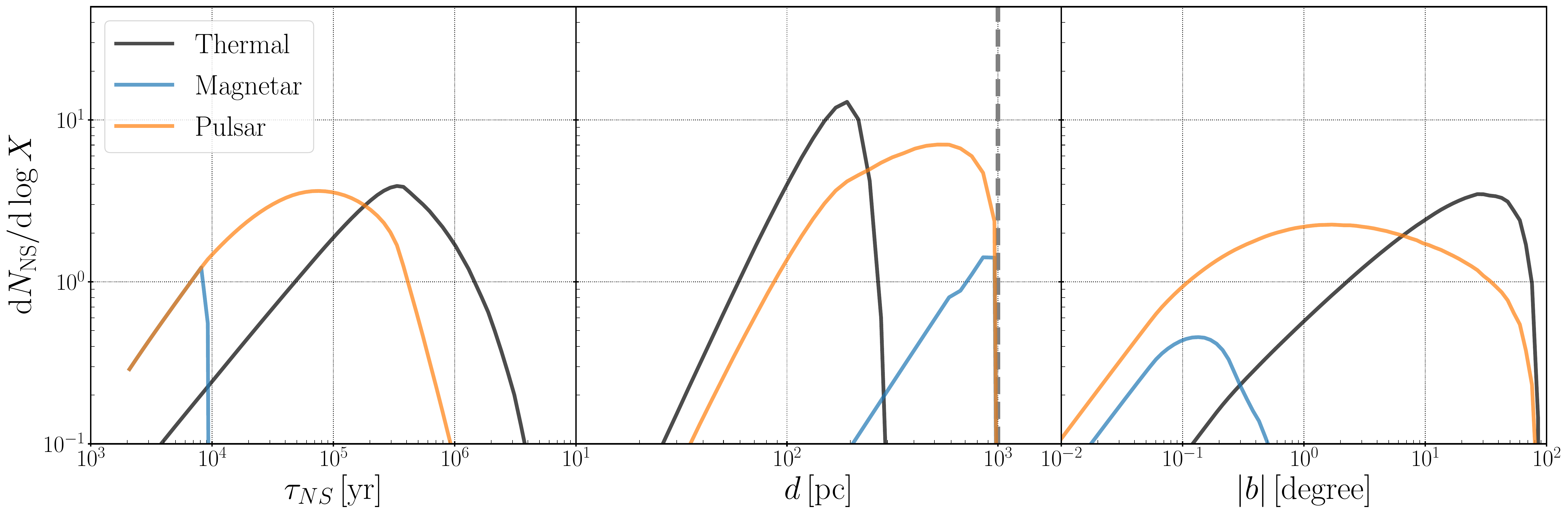}
\end{center}
\caption{Left, middle, and right panels show the number distribution of detectbale HVNSs as a function of age, distance from the Sun, and Galactic latitude, respectively. Black, orange, and blue lines represent the distribution of thermally emitting NSs ({\it model 4}), pulsars ({\it model 19}), and magnetars ({\it model 26}), respectively.}
\label{fig:hist}
\end{figure*}

With Eq.~(\ref{eq:number}), we estimate the number of detectable HVNSs with different parameter sets.
Setups and results of our model calculations are summarized in Tables~\ref{table:model1}-\ref{table:model3}.
To highlight the contribution of thermal and spin-down radiation to NS detectability, 
the models, where thermal and spin-down radiation are separately considered, are summarized in Tables~\ref{table:model1} and \ref{table:model2}, respectively.
The cases supposing the composite spectra are presented in Table~\ref{table:model3}.
For convenience, we hereafter refer to our calculation results with the model ID given in the first column of these tables.

First, we argue the feasibility of hunting thermally emitting NSs with {\it model 4}, supposing the low-mass NS cooling model and $\vkick = 500~\kms$.
This model calculation predicts that about 4 thermally emitting NSs are observable in the all sky.
This result is in good agreement with a simple order-estimate: when considering that we observe HVNSs with $\tauns \leq 10^6~\yr$ within the observable radius of $\robs \sim 200$ pc, the putative number is described as $\nns \sim \pi \robs^2~\sgmns(\rsun)~\tauns \sim 4.4$.
To further understand the properties of this solution, we show the distribution of detectable NSs in age, distance from the Sun, and the galactic latitude in Figure~\ref{fig:hist}.
While below $\tauns \sim 10^5$ yr the detectable number of thermally emitting NSs increases with age, it declines for older populations reflecting the rapid decrease of $\tns$ due to the NS cooling.
Similarly, the number of NSs rises as $\nns \propto d^2$ and sharply drops around 300 pc, corresponding to the observable radius for $\tauns \lesssim 10^5$ yr.
In this case, the detectable NSs are widely distributed beyond $|b| > 10^\circ$ because most of them are old enough to travel the distance of $\vkick \tauns \gtrsim 100$ pc and reach such a high-latitude region.

Next, we focus on the results of {\it model 19} and {\it 26}, which are the counterpart cases of {\it model 4} considering the spin-down radiation instead of the thermal one.
{\it Model 19} shows that about 5 pulsars are observable, which is almost comparable to the thermally emitting NS case.
For the detectable pulsars, the age distribution shows a peak at younger age than the thermally emitting NSs, since the pulsar spin-down radiation outshines the black body radiation until decaying around $\tauns \sim 10^5$ yr (see Figure~\ref{fig:luminosity}).
Consequently, the pulsars are further detectable up to $\rmax =$ 1 kpc, which is set for accurate measurements of proper motions by the LSST.

On the other hand, {\it model 26} predicts that the expected number of detectable magnetars is less than unity.
This result is simply due to the short lifetime of $\tauns \sim 10^4$ yr and the cutoff at $\rmax$.
We also note that due to the young age distribution, magnetars are still buried in the Galactic disk region suffering from significant dust extinction, which makes their detection more challenging.
To consider this observational circumstances, we evaluate the number of detectable HVNSs at $|b| > 5^\circ$.
{\it Models 4} and {\it 19} predict about 3 thermally emitting NSs and 1 pulsars at such high latitude, respectively, but almost zero magnetars in {\it model 26}.
Thus, it is desperate to detect magnetar-like HVNSs in the future optical observations by the LSST.

\begin{figure}
\begin{center}
\includegraphics[width=\columnwidth]{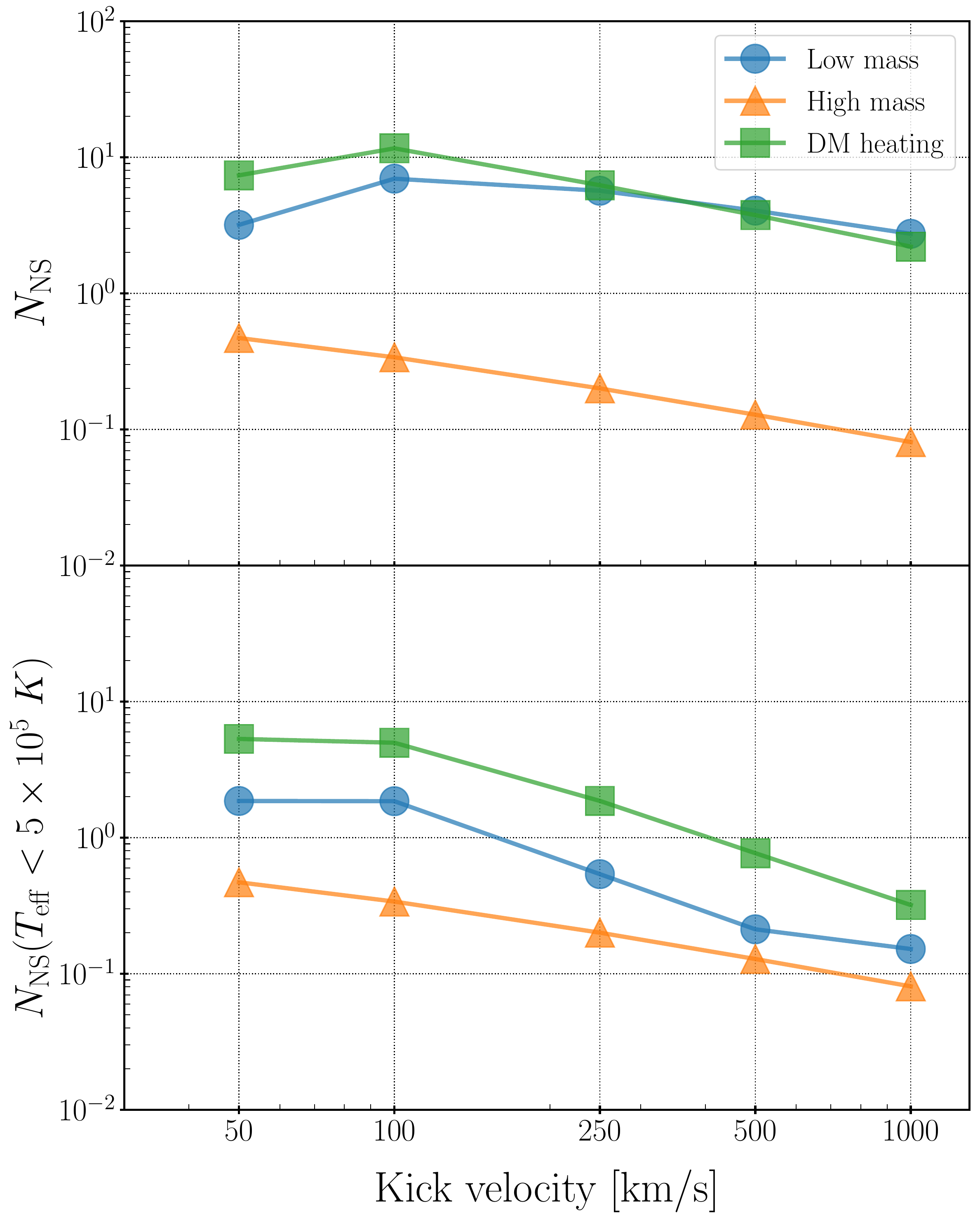}
\end{center}
\caption{Upper panel shows the number of detectable thermally emitting NSs as a function of kick velocities.
Cyan circles, orange triangles, and green squares represent the results calculated with Low mass, High mass, and DM heating models, respectively.
Lower panel is the same figure as upper one, but focusing on for low temperature NSs with $\tns < 5 \times 10^5$ K.
The data plotted here are also shown in Table~\ref{table:model1}.}
\label{fig:Nns}
\end{figure}

\subsection{Dependence of Model Parameters}\label{sec:dependence}

Here, we investigate the dependence of several parameters with relatively big uncertainties on our model calculation.
{\it Models~1-15} in Table~\ref{table:model1} compare the detectability of thermally emitting NSs between three NS cooling models; Low-mass, High-mass, and DM heating.
With these model calculations, we also demonstrate the effect of NS kick-velocity, which can exhibit a wide variety depending on properties of progenitors, such as mass, radii, and direction of supernova mass ejection.
Based on proper motions of observed radio pulsars, \citet{Hobbs2005MNRAS} has suggested the mean of NS kick-velocity to be $400~\pm 40~\kms$ and somewhat a large velocity dispersion $265~\kms$ \citep[see also][for a suggestion of better measurements with future astrometric observations of radio pulsars]{Ronchi2021arXiv}.
 The transverse velocities of magnetars are distributed from $100$--$350\,{\rm km/s}$ \citep[][and references therein]{Enoto_2019}.
We, therefore, assume five kick velocities of $\vkick =$ 50, 100, 250, 500 and 1000$~\kms$ for each NS cooling model.

Figure~\ref{fig:Nns} summarizes the results for {\it models 1-15}.
High mass cooling model predicts that the expected number of detectable thermally emitting NSs is always less than unity reflecting the low effective temperature and faint brightness. 
Thus, if most of NSs have efficiently cooled down as predicted by High mass model, hunting thermally emitting NSs would be challenging in the optical observation.
On the other hand, for Low mass cooling and DM heating models that relatively well reproduce the observed age-temperature relation of NSs, at least 2 thermally emitting NSs are detectable.
These two NS cooling models show a similar dependence of kick velocities on NS detection rates; $\nns$ reaches a peak at $\vkick = 100~\kms$ and gradually decreases for higher kick velocities.
This is because, with large kick velocities, NSs that originally form at the solar neighborhood can quickly escape from the observable radius, e.g., considering $\tauns = 10^6~\yr$, $\vkick \tauns > \robs$ at $\vkick \gtrsim 200~\kms$.

Here, it is worth predicting the number of detectable NSs with $\tns < 5 \times 10^5$ K, where optical observations would give unique NS samples that have been dimmed in X-ray band.
Figure~\ref{fig:Nns} shows that the expected number of such low-temperature NSs is 1.9, 0.5, and 5.3 for Low mass, High mass, and DM heating models, respectively, at $\vkick = 50~\kms$, and declines for higher kick velocities.
Note that the negative dependence of $\nns(\tns < 5 \times 10^5~\kelvin)$ on kick velocities is stronger than for the case of whole detectable NSs, since lower kick velocities allow more old stars to stay within the observable radii, as mentioned above.
In this case, DM heating model always predicts more detectable NSs than Low mass model, reflecting the high effective temperature of old NSs that is sustained by DM annihilation in cores. More importantly, the fraction of cold NSs ($T\lesssim 5\times 10^{5}\,{\rm K}$) is very sensitive to the cooling curve and the typical kick velocity.
Thus, the future detection of thermally emitting NSs especially with $\tns < 5 \times 10^5$ K will provide an essential information on NS thermal evolution.

The detectability of pulsars ({\it models 16-20}) and magnetars ({\it 23-27}) also depend on the kick velocities.
In contrast to the thermally emitting NSs, the detection rates of pulsars and magnetars increase with higher kick velocities.
This is because the observable volume of pulsars and magnetars is given by $\rmax$, limited by the observation of proper motions rather than the apparent magnitude, and thus, it shrinks for lower kick velocities.

The detectability of pulsars assuming different decay timescales of magnetic fields, 
is also examined. 
Practically this timescale is constrained to be longer than $10^7$ yr from the population arguments of luminous pulsars.
{\it Models 19, 21,} and {\it 22} provide no significant difference in pulsar detection within the range of $\tau_{\rm B} = 10^7$-$10^9$ yr.
This result naturally comes from the fact that, among these models, the electromagnetic signature of pulsars are identical untill $\tauns \sim 10^7$ yr, and most of detectable pulsars are much younger than this age.
Thus, the pulsar detection rates in the future optical observation is completely independent of the magnetic field decay timescale as long as $\tau_{\rm B} > 10^7$ yr.

Finally, we present the model prediction supposing the composite spectra of the thermal and spin-down radiation with {\it models 28-33} summarized in Table~\ref{table:model3}.
In these models, a half of NSs is assumed to be born as magnetars with $\tau_B = 10^4$ yr, and another half as pulsars with $\tau_B = 10^7$ yr.
DM heating model, the most optimistic case, predicts that about 12 (10) HVNSs are observable in the whole sky, and about 9 (4) at $|z| > 5^\circ$ for the case of $\vkick = 100$ (500) $\kms$, and even High mass model, the most pessimistic one, ensures more than one detection.
Moreover, for most of the cases, we expect non-zero detection of NSs with $\tns < 5 \times 10^5$ K, which would be too faint to be observed in X-ray band, and thus future optical observations by LSST can uniquely provide such low-temperature NS samples that are helpful for understanding the NS thermal evolution.
Note here that while High-mass model predicts the detection of such low-temperature NSs, these samples should be dominantly powered by spin-down radiation, so that their effective temperature cannot be measured straightforwardly.
Thus, the fraction of NSs dominated by thermal radiation in the whole sample gives an important implication on NS radiative models.


\section{Summary and Discussion}\label{sec:summary}

In this paper, we have studied the feasibility of hunting high-velocity kicked neutron stars in future optical observations.
NSs are generally faint in optical-bands, but can be detectable when they travel to the solar neighborhood.
Modeling the optical luminosity of NS, we have found the observable radii to be $100$--$1000\,{\rm pc}$ in LSST observations, depending on the NS cooling and spin-down mechanisms. Although it is expected that thermally emitting NSs with $T\gtrsim 5\times 10^{5}\,{\rm K}$ and pulsars respectively are
more efficiently searched in the X-ray band by  {\it eROSITA} on the SRG observatory \citep{Pires2017AN,Sunyaev2021} and by
 the radio surveys \citep[see, e.g.,][for the prospects with the SKA]{Smits2009A&A},  
the optical survey will provide a unique opportunity to study cold isolated NSs with $T\lesssim 5\times 10^{5}\,{\rm K}$.

We expect that about 10 HVNSs are detectable in the all sky, and the main candidates are pulsars with $\tauns \sim 10^4$--$10^5~\yr$ and thermally emitting NSs with $\tauns \sim 10^5$--$10^6~\yr$.
In this case, the detection of the magnetars, of which lifetime is quite short, would be challenging because they still reside near the Galactic disk plane, along which the observable radii are significantly reduced due to severe dust extinction.
Taking into account this observational limitation, we would observe about 3 thermally emitting NSs and 1 pulsars at $|b| > 5^\circ$.

The number of detectable thermally emitting NSs can reflect the thermal evolution of NSs.
A rapid cooling model, where effective temperature of NSs quickly declines below $10^5$ K by $\tauns \sim 10^5$ yr, predicts less than one detectable thermally emitting NS.
On the other hand, with less efficient cooling cases that generally reproduce the temperature-age relations observed for the isolated thermally emitting X-ray NSs, about 4 thermally emitting NSs are detectable.
Moreover, our calculations have suggested that NS heating mechanism via DM annihilation in cores at $\tauns > 10^6$ yr enhances the detection number of NSs with $\tns < 5\times 10^5$ K by a few factors, expected to be newly detected by the future optical observations.

In addition to NS cooling models, the distribution of NS kick velocities is an important factor in the detectability.
We found that the detection number has a peak of about 12 at $\vkick = 100~\kms$ and declines to about 2 at $1000~\kms$ for the case supposing DM heating in NS cores.
This result is naturally understood as NSs traveling more slowly can stay within the observable radius for longer time, and thus lower kick velocities allow more thermally emitting NSs to be observable.

We also have explored the dependence of $\vkick$ on the feasibility of hunting pulsars.
In contrast to the thermally emitting NSs, more detectable pulsars are predicted for higher kick velocities.
This is because pulsars are detectable much further than thermally emitting NSs reflecting their huge spin-down luminosity, and therefore, their observable volume is limited by the detectability of the proper motions rather than the observed flux.
Thus, the ratio of the number of detectable NSs powered by thermal and spin-down radiation depends on their underlying velocity distribution.

The observed  thermally emitting isolated NSs show the optical excess, which is typically a factor of $\sim 5$ -- $10$ in flux expected from the extrapolation of the X-ray flux \citep{Kaplan2011ApJ}. Notably, RX J2143.0+0654 has among the largest excess of a factor of more than $50$ at $5000~\mbox{\AA}$. If thermally-emitting isolated NSs  always  have such an excess, we expect that the number of detectable objects with the LSST increases from our estimates by a factor of $\sim 10$--$30$.
In addition, there are some objects, e.g., the magnetar 4U 01421+61 and RXJ0806.4-4123, that are known to exhibit a near infrared (NIR) excess possibly attributed to a debris disk  \citep{Wang2006Natur,Posselt2018ApJ}.  
Therefore, the NIR survey by the Roman Space Telescope would also be promising to find isolated NSs in the strategy presented in this work.

In summary, we conclude that the future optical wide-area surveys
will give us opportunities to discover optical  HVNS samples. The number of 
observed objects as a function of age, effective temperature, and proper motions will provide great implications for the nature of NS formation and thermal evolution of NSs.


\section*{Acknowledgements}
We thank K.~Hamaguchi, N.~Nagata, H.~Noda, T.~Qiu and K.~Kashiyama for useful discussions. 
This work was supported in part by the World Premier International
Research Center Initiative (WPI Initiative), MEXT, Japan, Reischauer Institute of Japanese Studies at Harvard University, JSPS
KAKENHI Grant Numbers 
JP15H05887, JP15H05893, JP15H05896, JP15K21733, JP18H04350, JP18H04358, JP19H00677, JP20H05850, and JP20H05855,
and JSPS Early-Career Scientists Grant Number 20K14513.



\bibliographystyle{mnras}
\bibliography{refs.bib} 



\bsp	
\label{lastpage}
\end{document}